\documentclass[aps,prd,twocolumn,showpacs,floats,floatfix,nofootinbib]{revtex4}
\usepackage{graphicx}
\usepackage{dcolumn}
\usepackage{bm}
\usepackage{graphics}
\usepackage{amssymb}
\usepackage{amsmath}
\usepackage{natbib}
\usepackage{amsmath}
\usepackage{url}
\usepackage{color}
\newcommand{\ba}{\begin{eqnarray}}
\newcommand{\ea}{\end{eqnarray}}
\newcommand{\dd}{{\rm d}}

\begin{document}
\title{On a Geometrical Description of Quantum Mechanics}

\author{M. Novello\footnote{M. Novello is Cesare Lattes ICRANet Professor}}
\email{novello@cbpf.br}
\author{J. M. Salim}
\email{jsalim@cbpf.br}
\author{F. T. Falciano}
\email{ftovar@cbpf.br}
\affiliation{Instituto de Cosmologia Relatividade Astrofisica ICRA - CBPF\\
 Rua Dr. Xavier Sigaud, 150, CEP 22290-180, Rio de Janeiro, Brazil}
\begin{abstract}
We show that Quantum Mechanics can be interpreted as a modification
of the Euclidean nature of 3-d space into a particular Weyl
affine space which we call Q-wis. This is proved using the Bohm-de
Broglie causal formulation of Quantum Mechanics. In the Q-wis
geometry, the length of extended objects changes from point to
point.  In our proposed geometrical formulation, deformation of the
standard rulers used to measure physical distances are in the core
of quantum effects.
\end{abstract}


\maketitle

\section{Introduction}

The early years of quantum mechanics were marked by intense
debates and controversies related to the meaning of the new behavior of matter.
While one group was convinced that was unavoidable to abandon the classical
picture, the other group tried incessantly to save its main roots and
conceptual pillars. To be able to reproduce the atypical quantum effects,
the latter group was forced to introduce new ingredients such as
de Broglie's pilot wave \cite{debroglie} or Mandelung's hydrodynamical
picture \cite{mandelung}.

However, the lack of physical explanations for these ad hoc
modifications weakened these pictures. At the same time, the former
group leaded by Schr\"odinger, Bohr and Heisenberg was increasingly
gaining new adepts until its climax in the 1927 Solvay's conference
when this picture was finally accepted as the orthodox
interpretation of quantum mechanics - the Copenhagen interpretation
\cite{copenhagen}.

Notwithstanding, a marginal group of physicists continued to develop
other approaches \cite{others} to describe quantum mechanics that
are more adequate to connect to a classical picture\footnote{Since
we are not concerned with relativistic phenomena, the term classical
physics should be understood as pre-relativistic physics unless
otherwise specified.}. One of the most prominent amongst these
alternative interpretations is the causal interpretation of quantum
mechanics also known as Bohm-de Broglie interpretation \cite{bdb}.

The development of quantum cosmological scenario brought to light
some difficulties intrinsic to the Copenhagen interpretation. More
specifically, the measurement process in a quantum closed universe
seems inevitably inconsistent \cite{nelson}. Fortunately, there are
some alternative interpretations that are consistently applied
simultaneously to  cosmology and to the micro-world. As two examples
we mention the many-worlds interpretation \cite{others2} and the
consistent histories formulation \cite{others3}.

In the present work we will focus only on the Bohm-de Broglie
interpretation since it is amongst the well defined interpretation
that can be applied to any kind of system, including the universe as
a whole, and up to date it is completely equivalent to the
Copenhagen interpretation when applied to the micro-world.

We will show that it is possible to interpret all quantum phenomena
as a modification of the geometrical properties of the physical
space. Hence, we will deal with a generalization of Euclidean
geometry that was first introduced by Weyl \cite{Weyl}.

Weyl proposed a different modification of the Euclidean geometry
than the one developed by Riemann. In fact, a Riemannian geometry
can be understood as a special case of a Weyl geometry. In appendix
\ref{qwis} we describe in more details the geometrical properties of
a Weyl geometry, but it is worth to mention its main difference that
is related to the notion of a standard ruler.

In a Weyl geometry the length of an extended object changes from
point to point. This means that a ruler of length $l$ will change by
an amount
\[
\delta \, l = l \, f_{a} \, \dd x^{a} \quad .
\]
This effect may become an obstacle to the notion of a local ruler
and thus to local measurement of distance. However, there is a
special class of Weyl geometries known as Weyl Integrable Space
(Q-wis) that is free of such difficulty.

This is provided by the condition that the vector $f_{i}$ is a
gradient of a function, i.e. $f_a=f_{,\, a}.$  Q-wis is
distinguished precisely by the fact that the length of the ruler
transported along a closed curve does not change. Hence, if the
change of the ruler's length is $\dd l$, for a closed path in Q-wis
we have
\[
\oint \dd l=0 \qquad ,
\]
which guarantees the uniqueness of any local measurement. The
allowance of an intrinsic modification of the standard rulers is
the main geometrical hypothesis of the present work.  Similarly to
London \cite{london} and Santamato \cite{santamato}, we shall argue
how this geometrical modification can be in the origin of quantum
effects. For the sake of clarity we will deal with the simplest
system possible, namely an isolated point-like particle possibly
subjected to an external potential.

The outline of the article is as follows. In the next section we briefly review the main points of quantum mechanics and in section \ref{NEG} we describe how to connect the Q-wis space to quantum theory. We show that quantum mechanics can be derived from a geometrical variational principle. In section \ref{conclusion} we present our final remarks. Appendix \ref{qwis} is reserved to describe the main properties of the Q-wis geometry.

\section{Quantum Mechanics}\label{QM}

Quantum mechanics is a modification of the classical laws of physics to
incorporate the uncontrolled disturbance caused by the macroscopic
apparatus necessary to realize any kind of measurement. This
statement, known as Bohr's complementary principle, contains the
main idea of the Copenhagen interpretation of quantum mechanics. The
quantization program continues with the correspondence principle
promoting the classical variables into operators and the Poisson
brackets into commutation relations.

In this non-relativistic scenario, the Schr\"odinger equation
establishes the dynamics for the wave function describing the
system. Note that as in newtonian mechanics time is only a external
parameter and the 3-d space is assumed to be endowed with the
Euclidean geometry.

Using the polar form for the wave function, $\Psi=A\, e^{iS/\hbar}$,
the Schr\"odinger equation can be decomposed in two equations for
the real functions $A\left(x\right)$ and $S\left(x\right)$
\begin{eqnarray}
&&\frac{\partial S}{\partial t}+\frac{1}{2m}\nabla S.\nabla S +V
-\frac{\hbar^{2}}{2m}\frac{\nabla^{2}A}{A}=0 \quad ,\quad \label{HJ}\\
&&\frac{\partial A^2}{\partial t}+\nabla \left(A^2\frac{\nabla S}{m}\right)=0 \qquad .\label{cont}
\end{eqnarray}

Solving these two equations is completely analogous to solving the Schr\"odinger
equation. The probabilistic interpretation of quantum mechanics
associate $\| \Psi \|^2=A^2$ with the probability distribution
function on configuration space. Hence, eq. (\ref{cont}) has exactly
the form of a continuity equation with $A^2 \nabla S/m$ playing the role
of current density.\\

\subsection{Bohm-de Broglie interpretation}

The causal interpretation which is an ontological hidden variable
formulation of quantum mechanics, propose that the wave function
does not contain all the information about the system.

An isolated system describing a free particle (or a particle
subjected to a potential $V$) is defined simultaneously by a wave
function and a point-like particle. In this case, the wave function
still satisfies the Schr\"odinger equation but it should also works as
a guiding wave modifying the particle's trajectory.

Note that eq.(\ref{HJ}) is a Hamilton-Jacobi like equation with an
extra term that it is often called quantum potential
\begin{equation}\label{Q}
Q=-\frac{\hbar^{2}}{2m}\frac{\nabla^{2}A}{A}\quad ,
\end{equation}
while, as already mentioned, eq.(\ref{cont}) is a continuity-like
equation. The Bohm-de Broglie interpretation take these analogies
seriously and postulate an extra equation associating the velocity
of the point-like particle with the gradient of the phase of the
wave function. Hence,
\begin{equation}\label{guide}
\dot{x}=\frac{1}{m}\nabla S \qquad .
\end{equation}

Integrating eq. (\ref{guide}) yields the quantum bohmian
trajectories. The unknown or hidden variables are the initial
positions necessary to fix the constant of integration of the above
equation.

The quantum potential is the sole responsible for all novelties of
quantum effects such as non-locality or tunneling processes. As a
matter of fact, the Bohm-de Broglie interpretation has the
theoretical advantage of having a well formulated classical limit.
Classical behavior is obtained as soon as the quantum potential,
which has dimensions of energy, becomes negligible compared to other
energy scales of the system.

In what follows, we will show that it is possible to reinterpret
quantum mechanics as a manifestation of non-Euclidean structure of
the 3-dimensional space. Hence, we propose a geometrical
interpretation to describe quantum effects.

\section{Non-Euclidean geometry}\label{NEG}

Since ancient times, Euclidean geometry was considered as the most
adequate mathematical formulation to describe the physical space.
However, its validity can only be established a posteriori as long
as its construction yields useful notions to connect physical
quantities such as the Euclidean distance between two given points.

Special relativity modified the notion of 3-dimensional Euclidean
space to incorporate time in a four-dimensional continuum (Minkowski
spacetime). Later on, General Relativity generalized the absolute
Minkowski spacetime to describe gravitational phenomena. General
Relativity considers the spacetime manifold as a dynamical field
that can be deformed and stretched but in such a way that it always
preserves its Riemannian structure. It is worth noting that both the
Euclidean and Minkowskian spaces are nothing more than special cases
of Riemaniann spaces.

Nonetheless, Riemannian manifold are not the most general type of
geometrical spaces. In the same way as above, Riemannian geometries
can be understood as a special subclass of a more general structure
known as Weyl space. As to the matter of which geometry is actually
realized in Nature, it has to be determined by physical experiments.

Instead of imposing a priori that quantum mechanics has to be
constructed over an Euclidean background as it is traditionally done,
we shall argue that quantum effects can be interpreted as a
manifestation of a non-Euclidean structure derived from a
variational principle. The validity of the specific geometrical
structure proposed can be checked a posteriori comparing it to the
usual non-relativistic quantum mechanics.

Thus, consider a point-like particle with velocity $v=\nabla S/m$
and subjected to a potential $V$. Following Einstein's idea to
derive the geometrical structure of space from a variational principle by
considering the connection as an independent variable, we start with
\begin{equation}\label{I}
I=\int{ \dd t\dd^3x\sqrt{g}} \, \Omega^2 \left(\lambda^2\mathcal{R}-\frac{\partial S}{\partial t}-\mathcal{H}_m\right) \quad ,
\end{equation}
and consider the connection of the 3-d space $\Gamma^i_{j k}$, the
Hamilton's principal function $S$ and the scalar function $\Omega$
as our independent variables.

Each one of the terms in equation (\ref{I}) is understood as
follows: we are considering the line element in cartesian
coordinates given by
\[
\dd s^2=g_{ij}\dd x^i \dd x^j=\dd x^2+\dd y^2+\dd z^2
\]
with
\[
g={\rm det}\, g_{ij} \quad.
\]
The Ricci curvature tensor is defined in term of the connection through
\[
\mathcal{R}_{ij}=\Gamma^m_{ mi\, ,j}-\Gamma^m_{ij\, ,m}+\Gamma^l_{mi}\Gamma^m_{jl}-\Gamma^l_{ij}\Gamma^m_{lm}
\]
and its trace defines the curvature scalar $\mathcal{R} \equiv
g^{ij}\mathcal{R}_{ij}$ which has dimensions of inverse length
squared, $[\mathcal{R}]=L^{-2}$. The constant $\lambda^2$ has
dimension of energy times length
squared, $[\mathcal{\lambda}^2]=E.L^{2}$, and the $\frac{\partial
S}{\partial t}$ term is related to the particle's energy. In the
case of our point-like particle the matter hamiltonian is simply
\[
\mathcal{H}_m=\frac{1}{2m}\nabla S.\nabla S +V \qquad .
\]

>From equation (\ref{I}), variation of the
action $I$ with respect to the independent variables results
(see appendix \ref{qwis} for details)
\begin{eqnarray}
\delta \Gamma^i_{j k}:&\quad&g_{ij;k}=-4\left(\ln \Omega\right)_{,k}\, g_{ij} \qquad ,\qquad \qquad \qquad \quad \quad \label{weq}
\end{eqnarray}
where ``$;$'' denotes covariant derivative and a common ``$,$''
simple spatial derivative. Equation (\ref{weq}) characterize the
affine properties of the physical space. Hence, the variational
principle naturally defines a Weyl Integrable Space. Variation with
respect to $\Omega$ gives
\begin{eqnarray}
\, \delta \Omega:&\quad& \lambda^2\mathcal{R}=\frac{\partial S}{\partial t}+\frac{1}{2m}\nabla S.\nabla S+V \quad .\qquad \qquad \qquad  \label{HJ3}
\end{eqnarray}

The right-hand side of this equation has dimension of energy while
the curvature scalar has dimension of $[\mathcal{R}]=L^{-2}$.
Furthermore, apart from the particle's energy, the only extra
parameter of the system is the particle's mass $m$. Thus, there is
only one way to combine the unknown constant $\lambda^2$, which has
dimension of $[\lambda^2]=E.L^{2}$, with the particle's mass such as
to form a physical quantity. Multiplying them, we find a quantity
that has dimension of angular momentum squared
$[m.\lambda^2]=\hbar^2$.

In terms of the scalar function $\Omega$, the curvature scalar is
given by (see appendix \ref{qwis})
\begin{equation}\label{R}
\mathcal{R}=8\frac{\nabla^2 \Omega}{\Omega} \qquad .
\end{equation}
Hence, setting $\lambda^2=\hbar^2/16 m$, equation (\ref{HJ3}) becomes
\begin{eqnarray}
\, \delta \Omega:&\quad&\frac{\partial S}{\partial t}+\frac{1}{2m}\nabla S.\nabla S+V -\frac{\hbar^2}{2m}\frac{\nabla^{2}\Omega}{\Omega}=0 \quad ,\qquad  \label{HJ4}
\end{eqnarray}
Finally, varying the Hamilton's principal function $S$ we find
\begin{eqnarray}
\, \delta S:&\quad&\frac{\partial \Omega ^2}{\partial t}+\nabla \left(\Omega^2\frac{\nabla S}{m}\right)=0 \qquad .\label{cont3}
\end{eqnarray}

Equations (\ref{HJ4}) and (\ref{cont3}) are identical to equations
(\ref{HJ}) and (\ref{cont}) if we identify $\Omega=A$. Thus, the
``action'' of a point-like particle non-minimally coupled to
geometry given by
\begin{equation}
I=\int{ \dd t \dd^3x\sqrt{g}} \, \Omega^2 \left[\frac{\hbar^2}{16\, m}\mathcal{R}-\left(\frac{\partial S}{\partial t}+\mathcal{H}_m\right)\right] \quad ,
\end{equation}
exactly reproduce the Schr\"odinger equation and thus the quantum behavior.

The straightest way to compare this geometrical approach to the
common quantum theories is to relate it to the Bohm-de Broglie
interpretation\footnote{Up to date, all interpretation of quantum
mechanics are on equal footing. Thus, establishing the connection
with the causal interpretation automatically links this geometrical
interpretation with all others.}. Note that this formulation has the advantage of giving a physical
explanation of the appearance of the quantum potential, eq.
(\ref{Q}). In Q-wis, this term is simply the curvature scalar of the
Weyl integrable space. The inverse square root of the curvature
scalar defines a typical length $L_w$ (Weyl length) that can be used
to evaluate the strength of quantum effects
\[
L_w \equiv \frac{1}{\sqrt{\mathcal{R}_w}} \quad .
\]

As we have already mentioned, the classical limit of Bohm-de Broglie
interpretation is achieved when the quantum potential is negligible
compared to other energy scales of the system. In the scope of this
geometrical approach, the classical behavior is recovered when the
length defined by the Weyl curvature scalar is small compared to the
typical length scale of the system. Once the Weyl curvature becomes
non-negligible the system goes into a quantum regime.

\subsection{Geometrical uncertainty principle}\label{gup}

As long as we accept that quantum mechanics is a manifestation of a
non-Euclidean geometry, we are faced with the need of reinterpreting
geometrically all theoretical issues related to quantum effects. As
a first step, we derive the uncertainty principle as a break down of
the classical notion of a standard ruler.

It is well known amongst relativistic physicists that there is no
absolute notion of spatial distance in curved spacetime. However,
this is no longer true when there is an absolute newtonian time and
only the spatial manifold is allowed to be curved. In this case, it
is possible to define distance as the smallest length between two
given points calculated along geodesics in 3-d space. This is a
consistent definition since the 3-d space has a true metric in the
mathematical sense that its eigenvalues are all positives. However,
this definition does not encompass the classical definition of a
standard ruler.

Hence, we are unable to perform a classical measurement to distances
smaller than the Weyl curvature length. In other words, the size of
a measurement has to be bigger than the Weyl length
\begin{equation}\label{princ1}
\Delta L \geq L_w= \frac{1}{\sqrt{\mathcal{R}_w}} \quad .
\end{equation}

The quantum regime is extreme when the Weyl curvature term
dominates. Thus, from equations (\ref{R}) and (\ref{HJ4}) we have
\begin{equation}\label{princ2}
\mathcal{R}_w =2\left(\frac{2\Delta p}{\hbar}\right)^2-\frac{16m}{\hbar^2}\left(\frac{\partial S}{\partial t}-V\right) \leq 2\left(\frac{2\Delta p}{\hbar}\right)^2
\end{equation}
and finally combining equations (\ref{princ1}) and (\ref{princ2}) we obtain
\[
\Delta L . \Delta p\geq \frac{\hbar}{2\sqrt{2}}\qquad .
\]

We should emphasize that now the Heisenberg's uncertainty relation
has a pure geometrical meaning. Our argument closely resembles
Bohr's complementary principle inasmuch as the impossibility of
applying the classical definitions of measurements. However, we
strongly diverge with respect to the fundamental origin of the
physical limitation.

Bohr's complementary principle is based on the uncontrolled
interference of a classical apparatus of measurement. On the other
hand, we argue that the notion of a classical standard ruler breaks
down because its meaning is intrinsically dependent on the validity
of Euclidian geometry. Once it becomes necessary to include the Weyl
curvature, we are no longer able to perform a classical measurement
of distance.

There is another way to interpret the uncertainty principle. For a
given particle of mass $m$ and energy $E$ there is only one
combination with the free parameter of the theory $(\hbar)$ that
furnishes a quantity with dimensions of length. We take this value
as a definition of the classical size of the particle, namely
\begin{equation}\label{size}
l_{part} \equiv \sqrt{\frac{\hbar^2}{E \, m}} \qquad .
\end{equation}

Note that this definition coincides with the Compton's wavelength of
the particle which is related to the limits of validity of
non-relativistic quantum mechanics.

Considering a free stationary particle, from equation (\ref{HJ3}) we
have
\[
E=\frac{\hbar ^2}{16m}\mathcal{R}_W  \; \Rightarrow \; l_{part} =\frac{4}{\sqrt{\mathcal{R}_W}} \qquad ,
\]
and from equation (\ref{princ2})
\begin{equation}
l_{part}\, . \Delta p \geq \sqrt{2}\, \hbar \qquad .
\end{equation}

>From this point of view, the uncertainty principle indicates that it
is impossible to perform a measurement smaller than the classical
size of the particle defined by equation (\ref{size}). In other
words, it is impossible to perform a classical measurement inside
the particle.

\section{Conclusions}\label{conclusion}

It is well known that as soon as we consider high velocities or high
energies one has to abandon the Euclidean geometry as a good
description of the physical space. These brought two completely
different modifications where the physical space loses its absolute
and universal character. In fact, these are the core of classical
relativistic physical theories, namely Special and General
Relativity.

In a similar way, one should be allowed to consider that the
difficulties that appears while going from classical to quantum
mechanics comes from an inappropriate extrapolation of the Euclidean
geometry to the micro-world. Hence, the unquestioned hypothesis of
the validity of the 3-d Euclidean geometry to all length scales
might be intrinsically related to quantum effects.

In the present work, we have shown that there is a close connection
between the Bohm-de Broglie interpretation of quantum mechanics and
the Q-wis Weyl integrable space. In fact, we point out that the
Bohmian quantum potential can be identified with the curvature scalar
of the Q-wis. Moreover, we present a variational principle that
reproduces the Bohmian dynamical equations considered up to date as
equivalent to Schr\"odinger's quantum mechanics.

The Palatini-like procedure, in which the connection acts as an
independent variable while varying the action, naturally endows the
space with the appropriate Q-wis structure. Thus, the Q-wis geometry
enters into the theory  less arbitrarily than the implicit ad hoc
Euclidean hypothesis of quantum mechanics.

The identification of the Q-wis curvature scalar as the ultimate origin of quantum effects leads to a geometrical version of the uncertainty principle. This geometrical description considers the uncertainty principle as a break down of the classical notion of standard rulers. Thus, it arises an identification of quantum effects to the length variation of the standard rulers.

\section*{Acknowledgments}

We would like to thank CNPq of Brazil and MN also to FAPERJ for financial support.
We would also like to thank the participants of the  ``Pequeno Semin\'ario'' of ICRA-CBPF's
Cosmology Group for useful discussions, comments and  suggestions. MN would like to
thank professor E. Elbaz for comments on this paper.

\appendix

\section{Q-wis Geometry}\label{qwis}


In this session we shall briefly review the mathematical properties of such 3-d Weyl Integrable Space (Q-wis).

Contrary to the Riemannian geometry which is completely specified by a metric tensor, the Weyl space defines an affine geometry. This means that the covariant derivative which is defined in terms of a connection $ \Gamma^{m}_{i k}$ depends not only on the metric coefficients but also on a vector field $f_a(x)$.

For instance, given a vector $ X_{a}$ its covariant derivative is
\begin{equation}
X_{a \, ; b} = X_{a \, , \, b} - \Gamma^{m}_{a b} \, X_{m} \qquad .
 \end{equation}

The non-metricity of the Weyl geometry implies that rulers, which
are standards of length measurement, changes while we transport it
by a small displacement $\dd x^{i}.$ This means that a ruler of
length $l$ will change by an amount
\begin{equation}
\delta \, l = l \, f_{a} \, \dd x^{a} \quad .
\end{equation}
As a consequence, the covariant derivative of the metric tensor does
not vanishes as in a Riemannian geometry but instead it is given by
\begin{equation}
g_{a b  \,; \, k} = f_{k} \, g_{a b }\qquad.
\end{equation}
Using cartesian coordinates, it follows that the
expression for the connection in terms of the vector $f_{k}$ takes
the form
\begin{equation}\label{conn}
\Gamma^{k}_{a b} = - \, \frac{1}{2} \, \left( \delta^{k}_{a} \,
f_{b} +  \delta^{k}_{b} \, f_{a} - g_{a b} \,  f^{k} \right) \quad.
 \end{equation}

The particular case of Weyl Integrable Space is provided by the
condition that the vector $f_{i}$ is a gradient of a function, i.e.
$f_a=f_{,\, a}$ . This property ensures that the length does not
changes its value along a closed path
\begin{equation}
\oint \, \dd l = 0 \quad.
\end{equation}

As a matter of convenience, we define
\begin{equation}
f = -4 \, ln \, \Omega \quad .
\end{equation}
Then the Ricci tensor
\[
\mathcal{R}_{ij}=\Gamma^m_{mi,j}-\Gamma^m_{ij,m}+\Gamma^l_{mi}\Gamma^m_{jl}-\Gamma^l_{ij}\Gamma^m_{lm}
\]
constructed with the above connection equation (\ref{conn}) is given by
\[
\mathcal{R}_{ij}=2\frac{\Omega_{,ij}}{\Omega}-6\frac{\Omega_{,i}\Omega_{,j}}{\Omega^2}+2g_{ij}\left[\frac{\nabla^2 \Omega}{\Omega} +\frac{\vec{\nabla} \Omega .\vec{\nabla} \Omega}{\Omega^2}\right]
\]
and the scalar of curvature $\mathcal{R}\equiv g^{ij}\mathcal{R}_{ij}$ becomes
\begin{equation}
\mathcal{R} =  \,8 \, \frac{\nabla^{2} \Omega}{\Omega} \quad.
\end{equation}

In the present paper we have used a variational principle to arrive at the Q-wis structure. The proof is as follows. Consider the action
\begin{equation}
I = \int \dd t\, \dd ^{3}x \sqrt{g} \; \, \Omega^{2} \, \mathcal{R}
\end{equation}
then, variation of the connection yields
\begin{eqnarray}
\delta I &=& \int \dd t\, \dd ^{3}x \sqrt{g} \; \, \Omega^{2}\, g^{ab} \, \delta R_{ab}
\nonumber \\
&=& \int \dd t \, \dd^{3}x\; Z^{a b}_{m}\, \delta \Gamma^{m}_{a b}
\end{eqnarray}
with
\begin{equation}\label{Zabm}
 Z^{a b}_{m} \equiv (\sqrt{g} \, g^{ab} \,
\Omega^{2})_{; \, m} - \frac{1}{2} \, (\sqrt{g} \, g^{a k} \, \Omega^{2})_{; \, k } \,
\delta^{b}_{m} - \frac{1}{2} \, (\sqrt{g} \, g^{b k} \, \Omega^{2})_{; \, k } \,
\delta^{a}_{m} \qquad
\end{equation}
Taking the trace of this expression yields
\begin{equation}\label{gakk}
\left(\sqrt{g} \, g^{a k} \, \Omega^{2}\right)_{; \, k } =0 \quad .
\end{equation}
Substituting (\ref{gakk}) in (\ref{Zabm}) we finally obtain the condition for a Weyl integrable geometry
\begin{equation}
 g_{a b \,; \, k} = - \, 4 \, \frac{\Omega_{, \, k}}{\Omega} \, g_{a b} \quad,
\end{equation}
or using that $g^{ik}g_{kj}=\delta ^i_{\, j}$ we find the contra-variant expression
\begin{equation}
 g^{a b}_{\phantom a \phantom a ; \, k} = 4 \, \frac{\Omega_{, \, k}}{\Omega} \, g^{a b} \quad .
\end{equation}


\end{document}